\documentstyle[12pt]{article}
\newread\epsffilein    
\newif\ifepsffileok    
\newif\ifepsfbbfound   
\newif\ifepsfverbose   
\newdimen\epsfxsize    
\newdimen\epsfysize    
\newdimen\epsftsize    
\newdimen\epsfrsize    
\newdimen\epsftmp      
\newdimen\pspoints     
\def\epsfbox#1{\global\def\epsfllx{72}\global\def\epsflly{72}%
   \global\def\epsfurx{540}\global\def\epsfury{720}%
   \def\lbracket{[}\def\testit{#1}\ifx\testit\lbracket
   \let\next=\epsfgetlitbb\else\let\next=\epsfnormal\fi\next{#1}}%
\def\epsfgetlitbb#1#2 #3 #4 #5]#6{\epsfgrab #2 #3 #4 #5 .\\%
   \epsfsetgraph{#6}}%
\def\epsfnormal#1{\epsfgetbb{#1}\epsfsetgraph{#1}}%
\def\epsfgetbb#1{%
\openin\epsffilein=#1
\ifeof\epsffilein\errmessage{I couldn't open #1, will ignore it}\else
   {\epsffileoktrue \chardef\other=12
    \def\do##1{\catcode`##1=\other}\dospecials \catcode`\ =10
    \loop
       \read\epsffilein to \epsffileline
       \ifeof\epsffilein\epsffileokfalse\else
          \expandafter\epsfaux\epsffileline:. \\%
       \fi
   \ifepsffileok\repeat
   \ifepsfbbfound\else
    \ifepsfverbose\message{No bounding box comment in #1; using defaults}\fi\fi
   }\closein\epsffilein\fi}%
\def\epsfsetgraph#1{%
   \epsfrsize=\epsfury\pspoints
   \advance\epsfrsize by-\epsflly\pspoints
   \epsftsize=\epsfurx\pspoints
   \advance\epsftsize by-\epsfllx\pspoints
   \epsfxsize\epsfsize\epsftsize\epsfrsize
   \ifnum\epsfxsize=0 \ifnum\epsfysize=0
      \epsfxsize=\epsftsize \epsfysize=\epsfrsize
     \else\epsftmp=\epsftsize \divide\epsftmp\epsfrsize
       \epsfxsize=\epsfysize \multiply\epsfxsize\epsftmp
       \multiply\epsftmp\epsfrsize \advance\epsftsize-\epsftmp
       \epsftmp=\epsfysize
       \loop \advance\epsftsize\epsftsize \divide\epsftmp 2
       \ifnum\epsftmp>0
          \ifnum\epsftsize<\epsfrsize\else
             \advance\epsftsize-\epsfrsize \advance\epsfxsize\epsftmp \fi
       \repeat
     \fi
   \else\epsftmp=\epsfrsize \divide\epsftmp\epsftsize
     \epsfysize=\epsfxsize \multiply\epsfysize\epsftmp
     \multiply\epsftmp\epsftsize \advance\epsfrsize-\epsftmp
     \epsftmp=\epsfxsize
     \loop \advance\epsfrsize\epsfrsize \divide\epsftmp 2
     \ifnum\epsftmp>0
        \ifnum\epsfrsize<\epsftsize\else
           \advance\epsfrsize-\epsftsize \advance\epsfysize\epsftmp \fi
     \repeat
   \fi
   \ifepsfverbose\message{#1: width=\the\epsfxsize, height=\the\epsfysize}\fi
   \epsftmp=10\epsfxsize \divide\epsftmp\pspoints
   \vbox to\epsfysize{\vfil\hbox to\epsfxsize{%
      \includegraphics{#1}%
      \hfil}}%
\epsfxsize=0pt\epsfysize=0pt}%

{\catcode`\%=12 \global\let\epsfpercent=
\long\def\epsfaux#1#2:#3\\{\ifx#1\epsfpercent
   \def\testit{#2}\ifx\testit\epsfbblit
      \epsfgrab #3 . . . \\%
      \epsffileokfalse
      \global\epsfbbfoundtrue
   \fi\else\ifx#1\par\else\epsffileokfalse\fi\fi}%
\def\epsfgrab #1 #2 #3 #4 #5\\{%
   \global\def\epsfllx{#1}\ifx\epsfllx\empty
      \epsfgrab #2 #3 #4 #5 .\\\else
   \global\def\epsflly{#2}%
   \global\def\epsfurx{#3}\global\def\epsfury{#4}\fi}%
\def\epsfsize#1#2{\epsfxsize}
\def\pixdir{}

\def\twofactor{.4}

\def\pixtype#1{\let\epsfinsert=\epsfbox[#1]}

\def\picture#1#2#3{\begin{tabular}{c}%
\mbox{\epsfxsize=#1\epsfinsert{\pixdir#2}}\\\mbox{#3}\end{tabular}}

\def\twoinrow#1#2#3#4{%
\mbox{\picture{\twofactor\hsize}{#1}{#2}\hspace{.05\hsize}%
\picture{\twofactor\hsize}{#3}{#4}}}

\def\fplot#1#2{\mbox{\epsfxsize=#1\hsize%
\epsfbox[65 10 782 535]{\pixdir#2}}}

\let\epsfinsert=\epsfbox
\textwidth=6.5in
\topmargin=-1in
\oddsidemargin=0in
\evensidemargin=0in
\textheight=9in

\begin{document}
\begin{titlepage}
\vbox{\hfill UM-P-92/90}
\begin{center}
\vspace{4cm}
{\large\bf Fermion Doubling and Gauge Invariance\\on\\Random Lattices}\\
\vspace{1cm}
{C. J. Griffin and T. D. Kieu,}\\
\vspace{1cm}
School of Physics,\\
University of Melbourne,\\
Parkville VIC 3052,\\
AUSTRALIA\\
\vspace{2cm}
\begin{quotation}
Random-lattice fermions have been shown to be free of the doubling problem
if there are no interactions or interactions of a non-gauge nature.
On the other hand, gauge interactions impose stringent constraints
as expressed by the Ward-Takahashi identities which could
revive the
free-field suppressed doubler modes in loop diagrams.  Comparing
random lattice, naive and Wilson fermions in two dimensional abelian background
gauge
theory, we show that indeed the doublers are revived for random lattices in the
continuum limit.  Some implications of the persistent doubling phenomenon
on random lattices are also discussed.
\end{quotation}
\end{center}
\vspace{5mm}{\small PACS: 11.15.Ha, 11.30.Rd}\hfill
\end{titlepage}
\topmargin=0in

\section{Introduction}
The doubling problem of lattice fermions is inevitable according to the
Nielsen Ninomiya no-go theorem\cite{NN} if the free-field action satisfies
the conditions of reflection positivity, locality, global axial symmetry,
and translational invariance at a fixed scale.
An obvious resolution of the doubling problem is thus to relax one of those
conditions to obtain, in the order listed above,
non-hermitian\cite{non-herm}, non-local\cite{non-local},
Wilson\cite{wilson}, or random-lattice\cite{ran,ran2,espiru,espiru2}
fermion formulations.
These formulations are all free of doublers when there are no
interactions or when the interactions are of a non-gauge
nature\cite{peran,tdkthesis}: the extra poles in the propagators
are removed as the lattice spacing $a$ decreases, leaving a single fermion mode
in the continuum limit.

Gauge interactions behave very differently on account of a
unique and special property.  Local gauge invariance imposes severe
constraints on the theory, expressed mathematically in the Ward-Takahashi
identities.  In particular, the fermion-gauge vertex is related to the
free inverse propagator,
\begin{eqnarray}
V_\mu(p) &\sim& \partial_\mu G_0(p),
\end{eqnarray}
giving the interaction vertices mode dependency.
This different coupling
strength of doublers to gauge fields has been
shown to revive these modes in loop diagrams, even though
they are suppressed at the  free-field level, in studies of some non-local
\cite{bodwin} and non-hermitian
formulations\cite{kashiwa,tdkprep}.
For this reason, we investigate the issue of fermion
doubling on random lattices with gauge interactions\cite{wheater}.


In the random lattice approach, suitable quantities are measured
on a random lattice then averaged, either quenchedly or annealedly,
over an ensemble of lattices.  Apart from the extra work involved in
generating an ensemble of random lattices, this approach better approximates
the scale-free rotational and translational symmetry of the
continuum than regular lattices.  Thus, the continuum limit
may be more easily reached on random lattices than on regular
lattices of the same size.  More relevant to this discussion,
since there is no
fixed Brillouin zone, there need be no extra poles of the
propagator. Even if extra poles do exist, the one-to-one correspondence between
propagator poles in momentum space and zero modes is not necessarily valid
since plane waves are
no longer eigenstates of the Dirac operator. Alternatively, one could appeal to
the fact that there is no transfer matrix on a random lattice (at least
for a finite lattice) since there are no identical timeslices, to
argue that there may not be a clear relation between poles of the
inverse propagator and the particle spectrum\cite{espiru2}.

This expectation of no doubling on random lattices has been realised
in various studies of free-field theory in both two and
four dimensions\cite{ran2,espiru}.  It has similarly been shown that random
lattice
theories with four-point interactions are also doubler free\cite{peran}.

\section{Our Approach}
We wish to compute the fermion determinant of abelian background gauge theory
on a
two dimensional Euclidean random lattice,
\begin{eqnarray}
-\ln{\rm
Det\,}(G_AG_0^{-1})&=&(G_A^{-1}G_0-1)-\frac{1}{2}(G_A^{-1}G_0-1)^2+O(g^4),
\end{eqnarray}
where $G_A^{-1}$ is the fermion propagator in the background gauge field
$A_\mu$.
Using background fields ensures
our results will not be marred
by internal gauge interactions, hence we expect to see a clean signal
which increases with the number of fermion flavours contributing to the
internal lines.
Comparing with identical calculations for naive and Wilson fermions on square
lattices,
which are known to be four-fold doubling and doubler-free respectively,
clarifies the continuum limit behaviour of our random lattices.

Our lattice is constructed from a triangulated array of $N$ square lattice
verticies
by a random sequence of Alexander ``flip'' moves, supplemented by
further constraints which force the lattice to stay
locally flat throughout the flipping procedure.
We chose to randomise the lattice with $N$ successful flips.
See figure \ref{fig.lat}.

\begin{figure}
\begin{center}
\twoinrow{unflip.ps}{initial lattice}{flip.ps}{randomised lattice}
\end{center}\caption{\label{fig.lat}A typical $8\times8$ lattice}
\end{figure}

This fixed vertex construction has several advantages over the lattices of
\cite{ran}: Construction is $O(N)$, compared with $O(N^3)$, and the resulting
lattice
has a fixed vertex spacing (a fixed size), so measured quantities do not need
to be adjusted for different link-lengths.
The trade off is that we have introduced a lattice dependent internal scale,
$s$, which will need to be dealt with.

The fermion derivative is chosen such that it reduces to the naive result
on lattices of regular arrangements of links. It is constructed by averaging
the contributions of pairs of consecutive links $(k,l)$ to the derivative in
lattice framing co-ordinates,
\begin{eqnarray}
\gamma_\mu\partial_\mu\Psi &\rightarrow &\frac{1}{C}\sum_{i=1\ldots
C}^{(k,l)_i}
\frac{\gamma}{k\times
l}\,\times\left[l\,\Psi(x+k)-k\,\Psi(x+l)+(k-l)\Psi(x)\right]
\end{eqnarray}
Gauge interactions are introduced in the usual gauge-invariant manner using the
link variables
$U_{x,x+l}=\exp(ig\int_lA(x)\,.\,{\rm d}x)$.
An alternative definition $U_{x,x+l}=\exp(ig\,l\,.\,A(x+l/2))$ which allows
gauge invariance to be explicitly broken on the lattice,
is also considered. The resulting action is hermitian, local and
axially-symmetric.

Measurements are made in a background gauge field
\begin{eqnarray}
g A_\mu&=&\delta_{\mu,1}\frac{E\sqrt{N}}{2\pi a}\sin\left(\frac{2\pi x_0}{a
\sqrt{N}}\right)
\end{eqnarray}
with fixed physical quantities
$ma^{-1}=0.1\ ({\rm length})^{-1}$, $a^2N=64\ ({\rm length})^{2}$, and
$a^{-2}E=0.05\ ({\rm length})^{-2}$, for $a=\{1.0,0.5,0.3333,0.25\}$
\section{Results}
\begin{figure}\begin{center}
\fplot{1}{fpg.ps}
\end{center}
\caption{\label{fig.fpg} Fermion propagation, $f(\xi)$.}
\end{figure}
We first compute a quantity derived from the free propagator
\begin{eqnarray}
f(\xi)&=&{\rm
Tr_\gamma}\,\frac{1}{N}\int_{x,x'}(1+\gamma_0)\,G_0^{-1}(x,x')\,\delta^1(x_0-x'_0+\xi),
\end{eqnarray}
evaluating the average zero momentum real particle propagator projected along
the $x_0$ direction. Figure
\ref{fig.fpg} summarises the calculation,
clearly identifying the doubler
suppression of free fermions on a random lattice in agreement with
\cite{espiru}. Indeed, apart from some minor small
distance fluctuations of the order of the internal scale,
the random lattice result matches the continuum
completely; the normalisation is reproduced exactly, and masses do not need to
be tuned%
{}.

The calculation of $-\ln{\rm Det\,}(G_AG_0^{-1})$ is complicated a little by
its sensitivity to
the internal scale of the lattice. Ideally we would like to ignore this effect
as
a small correction, or have
the internal scale match the lattice spacing $\langle s\rangle\sim a$. In
practice,
$\langle s\rangle\sim 1.3a$, the contribution is significant, and
$\langle s\rangle$ gets larger as the degree of randomisation is increased.
Using an ensemble of lattices, it is possible to extrapolate to
$\langle s\rangle=a$,
leading to a mean value of this extrapolation, and uncertainties associated
with the spread of geometrically different lattices that have the same $\langle
s\rangle$.
\begin{figure}\begin{center}
\fplot{1}{loop.ps}\end{center}
\caption{\label{fig.loop}Variation of $-\ln{\rm Det\,}(G_AG_0^{-1})$ with $a$}
\end{figure}

See figure \ref{fig.loop} for the results. The naive case
approaches the continuum limit quadratically with $a$ and the Wilson approaches
$\sim \frac{1}{4}$ of the same result linearly, as expected.
The same graph also indicates the random lattice results;
the number of lattice configurations used in the extrapolation is displayed
next to each point.
The lattice gauge invariant calculation is clearly more like the naive fermion
than the Wilson.
Moreover, had extrapolation in $\langle s\rangle$
not been performed, the result
would have been even larger. With gauge invariance broken, the converse is
clearly
seen; the result is certainly more like Wilson than naive, and in this case the
extrapolation procedure has forced an increase. Another amusing observation is
that the gauge non-invariant calculation approaches the continuum result
more rapidly than either Wilson or naive formulations.

\section{Discussion}
It is clear from our results that there are doublers on random
lattices when gauge invariance is maintained at finite lattice spacing,
since the extrapolated determinant is comparable to that of naive
fermions.  It can also be seen that the doubling can be avoided if one
gives up gauge invariance (but needs and hopes to recover it in the
continuum limit, as is in the case we studied above).

In all cases, the lattice fermion actions are invariant under the
global axial transformations.  When there are doublers on random lattices,
the axial anomalies are canceled in the usual manner among opposite-chirality
species.  When there is no doubling in the gauge non-invariant formulation,
the conserved lattice current being the Noether current of axial
symmetry is, of course, not gauge invariant.  Thus it cannot be
identified with the continuum axial current which is invariant.
It should be, instead, identified with a combination of the continuum
current and a gauge-noninvariant term, whose divergence gives us the
axial anomalies,
\begin{eqnarray}
J^{5\mu}_{\rm lattice}(x) &=& J^{5\mu}_{\rm continuum}(x) + \alpha
\epsilon^{\mu\nu}A_\nu(x).
\end{eqnarray}


We believe that the results obtained here are also applicable to other
kinds of random lattices.  Our doubling conclusion for random lattices is
not plainly disappointing but also points to some serious implications.

We have extended the lattice no-go theorem and at the same time
emphasised the importance of gauge invariance in the phenomenon of
lattice fermion doubling.

The failure of random lattices to accommodate chiral fermions could
either undermine the point of view that at the Planck scale or higher the
structure of spacetime is that of randomness; or, taken with other
complete failures in dealing with chiral fermions, could be a hint that
our understanding of chiral gauge theories is incomplete.  And,
correspondingly, the quantisation of those theories is in need of further
studies.  One of us has been pursuing this latter path\cite{tdkcgt}.

We acknowledge Dr Lloyd Hollenberg, Dr Girish Joshi and Prof Bruce
McKellar for their interests and support and Prof Tetsuyuki
Yukawa for discussions and guidance on random lattices.  TDK wishes to
thank Dr John Wheater and the Edinburgh Theory Group for discussions,
and acknowledges the support of an Australian Research
Council Fellowship and the
Pamela Todd Award.

\end{document}